\documentclass[a4paper,11pt]{article}
\usepackage[T1]{fontenc}
\usepackage[utf8]{inputenc}
\usepackage{amssymb}
\usepackage{amsmath}
\usepackage{amsfonts}
\usepackage{authblk}
\usepackage{graphicx}
\usepackage{hyperref}
\usepackage{cite}
\usepackage{setspace}
\usepackage{doi}
\begin{document}
\title{Nematic-isotropic transition in a density-functional theory for hard spheroidal colloids}
\author[1,2]{Eduardo dos S. Nascimento\thanks{\emph{Present address:} 
		Department of Physics, PUC-Rio, Rio de Janeiro, RJ, Brazil, \\
		\emph{e-mail}: edusantos18@esp.puc-rio.br}}
\affil[1]{Instituto de Física, USP, São Paulo, SP, Brazil}
\affil[2]{Liquid Crystal Institute, Kent State University, Kent, OH, USA}            
\date{} 
%
%
%
\maketitle
\begin{abstract}
We introduce a density-functional formalism based on the Parsons-Lee and the 
generalized van der Waals theories in order to describe the thermodynamics of 
anisotropic particle systems with steric interactions. For ellipsoids of 
revolution, the orientational distribution function is obtained by minimizing 
the free energy functional and the equations of state are determined. The system 
exhibits a nematic-isotropic discontinuous transition, characterized by a phase 
separation between nematic and isotropic phases at finite as well low packing 
fractions. The model presents a phase behavior which is in good agreement with 
Monte-Carlo simulations for finite aspect ratios.
\end{abstract}
%
%
%
%
%
%

%
%
\section{Introduction}

Anisotropic bodies interacting through steric potentials exhibit a complex phase behavior which is
the result solely of entropic effects \cite{Mederos,Dijkstra,Frenkel2014,Damasceno}. Many theoretical 
investigations have been developed in order to understand the macroscopic properties of non-spherical 
particle systems at equilibrium. Hard spherocylinders present isotropic fluid and solid phases, 
which are also expected to be stable in one-component hard spheres \cite{Tarazona}, in addition to 
liquid crystal phases, such as nematic and smectic structures \cite{Bolhuis}. Ellipsoids of revolution 
with steric interactions can give rise to isotropic, nematic and various solid phases 
\cite{FrenkelMulder1984,FrenkelMulder1985,Odriozola}, but there exist doubts if such system may have 
stable smectic phase \cite{Evans,Velasco}. Hard convex polyhedra may exhibit crystal structures, 
plastic crystals and liquid-crystalline phases depending on the anisotropic shape of particles \cite{Damasceno}.

Besides the theoretical interest, there are various experimental results on 
non-spherical bodies systems. Colloidal silica rodlike particles exhibit nematic 
and smectic phases depending on the volume fraction and aspect ratio\cite{Kuijk}. 
Suspensions of gibbsite platelets are found to present a nematic-isotropic phase 
coexistence as the volume fraction is varied \cite{Kooij}. Goethite nanorods have 
shown nematic mesophases with very interesting magnetic properties \cite{Lemaire}. 
These results suggest that mineral colloidal particles are promising materials for 
the study of liquid-crystalline phases \cite{Lekkerkerker}. 

One of the simplest model for colloidal anisotropic particles is the second-virial 
truncated free energy introduced by Onsager \cite{Onsager}. According to his 
calculations, the Helmholtz free energy can be written in the framework of a 
density-functional theory (DFT). For hard rodlike bodies, the model presents an 
entropy-driven nematic-isotropic (NI) discontinuous transition. However, Onsager's 
theory requires very low concentrations to describe quantitatively the phase behavior of 
the system. Also, particles should be very long in order justify the truncation of 
the virial series at the second virial coefficient \cite{Onsager,Straley}. Cleary, 
these assumptions can make comparison with the experiments and simulations a difficult task.

Many different DFT have been developed for systems of anisotropic particles with realistic dimensions 
and at finite volume fractions \cite{Mederos,Cinacchi,Parsons,Lee,Bolhuis,NascimentoPalffy}. A very successful 
proposal was presented by Parsons \cite{Parsons} and Lee \cite{Lee}, based on the idea that it is possible 
to decouple effectively the orientational and translational degrees of freedom. The Parsons-Lee (PL) theory 
leads to a configurational free energy which is a product of two terms, one is related to an hard-sphere 
(HS) reference system and other associated with the pair excluded volume. The HS system is usually 
described by the Carnahan-Starling (CS) theory \cite{CS}, which gives an analytical expression for 
the equation of state in a good agreement with virial expansions. For hard spherocylinders \cite{Lee} 
and hard spheroids \cite{Lee}, PL scheme gives a NI transition consistent with Monte-Carlo (MC) 
simulations even for short aspect ratios. Also, the theory recovers the Onsager's model \cite{Onsager} 
for very long rods at low densities. 

Although the CS equation of state is chosen to model the system of hard spheres, 
it is known that there exist many alternative proposals \cite{VeraSandler,Wei,Nord,Freasier} 
for dense isotropic fluids. A physically insightful approach is presented 
by Nordholm \textit{et al.} \cite{Nord} and Freasier \textit{et al.} \cite{Freasier}, 
which calculated the configurational free energy by using a generalized van der Waals 
(GvdW) approximation \cite{VeraSandler}. This theory argues heuristically that the 
partition function is associated with the free volume, which is the volume available 
to particles in the system. In fact, the free volume is estimated through an 
interpolation between the simple van der Waals (vdW) theory, which is appropriate 
for diluted systems, and a GvdW scheme constructed to model high density limit. 
Nordholm \textit{et al.} \cite{Nord} shown that, for a uniform HS fluid, the GvdW 
theory leads to an equation of state that is consistent with CS approximation at 
finite densities. Freasier \textit{et al.} \cite{Freasier} found good agreements 
in the oscillatory behavior of the HS radial distribution function calculated by 
means of GvdW scheme. These results indicates that the GvdW theory is a reasonable 
approximation in order to investigate systems of hard spheres at least qualitatively. 

We are interested in anisotropic fluid phases at finite as well high densities. 
Due to the importance and successful results, we do believe there is still room 
for studying the NI transition by means of a PL scheme. Specially, by means of 
alternative models for the reference system of hard spheres. In this work, we 
introduce a PL theory where the HS system is described by the GvdW theory 
developed by Nordholm \textit{et al.} \cite{Nord} and Freasier \textit{et al.} 
\cite{Freasier}. We write a simple expression for free energy functional in 
terms of the orientational distribution function. The equations of state are 
calculated exactly and the phase behavior is determined. Comparison with MC 
results \cite{Camp1996} shows that the model underestimates the 
NI transition for finite packing fractions. However, the theory agrees 
qualitatively well with the original PL theory and the simulation data 
for finite aspect ratios.

The paper is organized as follows. In Sec. \ref{FreeFunc} we introduce the free energy 
functional along the lines of PL and GvdW theories. Considering hard spheroids, the 
stationary distribution function is determined in Sec. \ref{StatDis}. The equations of 
state are calculated in Sec. \ref{EoS}, as well the phase behavior in terms of the packing 
fraction and the axial ratio. Finally, we present our conclusions in Sec. \ref{Con}.

\section{Free energy functional} \label{FreeFunc}

Consider a system of interacting rigid convex bodies, where $\boldsymbol{q}_{i}$ 
is the generalized coordinate of particle $i$, specifying the position of the 
center of mass and the orientation of the body-fixed frame. DFT formalism 
\cite{Mederos} establishes that we can write the configurational Helmholtz 
free energy as
\begin{equation} \label{Helden1}
 F [\varrho] = F_{id} [\varrho] + F_{ex} [\varrho],
\end{equation}
which is a functional of the number density $\varrho(\boldsymbol{q})$ such as
\begin{equation} \label{numdens}
 N = \int d \boldsymbol{q} \varrho(\boldsymbol{q}),  
\end{equation}
is the number of particles occupying the volume $V$, $F_{id}$ is the 
ideal part and $F_{ex}$ is the excess part, which presents all 
contributions related with particle interactions.

We are interested in spatially homogeneous phases in an ensemble 
of cylindrically symmetric bodies. Then, the local number density 
can be written as
\begin{equation}
 \varrho(\boldsymbol{q}) =\rho f(\hat{\boldsymbol{a}}),
\end{equation}
where $\rho=N/V$ and $f(\hat{\boldsymbol{a}})$ is the probability 
density of finding a particle with orientation given by $\hat{\boldsymbol{a}}$. 
Consequently, the ideal free energy is given by
\begin{equation} \label{Felid1}
  F_{id} = NkT \left[ \ln \rho - 1 + \int d\hat{\boldsymbol{a}} 
  f(\hat{\boldsymbol{a}}) \ln f(\hat{\boldsymbol{a}}) \right],
\end{equation}
which is related to the translational and orientational entropies. 
The main problem is to determine a reasonable good approximation 
for the excess free energy, since it is very difficult to
deal with the statistical mechanics of hard particle systems.

Here, we follow the PL theory \cite{Mederos,Parsons,Lee}, which 
basically states that the excess free energy can be approximated to
\begin{equation} \label{Felex1}
 F_{ex} = NkT \frac{ J\left( \eta \right)} {8v_0}
 \int d\hat{\boldsymbol{a}} d\hat{\boldsymbol{a}}^{\prime} f(\hat{\boldsymbol{a}}) 
 f(\hat{\boldsymbol{a}}^{\prime}) V_{e}\left( \hat{\boldsymbol{a}},\hat{\boldsymbol{a}}^{\prime} \right),
\end{equation}
where $J\left( \eta \right)$ is a function of the packing fraction 
$\eta=\rho v_{0}$, $v_0$ is the particle volume and $V_{e}\left( 
\hat{\boldsymbol{a}},\hat{\boldsymbol{a}}^{\prime} \right)$ is the 
excluded volume associated with a pair of particles oriented along 
$\hat{\boldsymbol{a}}$ and $\hat{\boldsymbol{a}}^{\prime}$, respectively. 
The quantity $J\left( \eta \right)$ should be chosen as the dimensionless 
free energy per particle of hard-sphere system. Usually, as Parsons 
\cite{Parsons} and Lee \cite{Lee} did, $J\left( \eta \right)$ is given by
\begin{equation} \label{CS}
 J^{\text{CS}} \left( \eta \right) = \frac{\left( 4-3\eta \right)\eta}{\left( 1-\eta \right)^2},
\end{equation}
which leads to the well-known CS equation of state \cite{Mederos,CS}. 
Clearly, that is the best choice to model a HS system. 
Adopting this approach, Lee shown \cite{Lee} that 
a system of hard ellipsoids of revolution presents a NI transition which agrees with 
MC simulations \cite{FrenkelMulder1984,FrenkelMulder1985} for finite densities 
and aspect ratios.

\begin{figure}
\centering
\includegraphics*[scale=0.3]{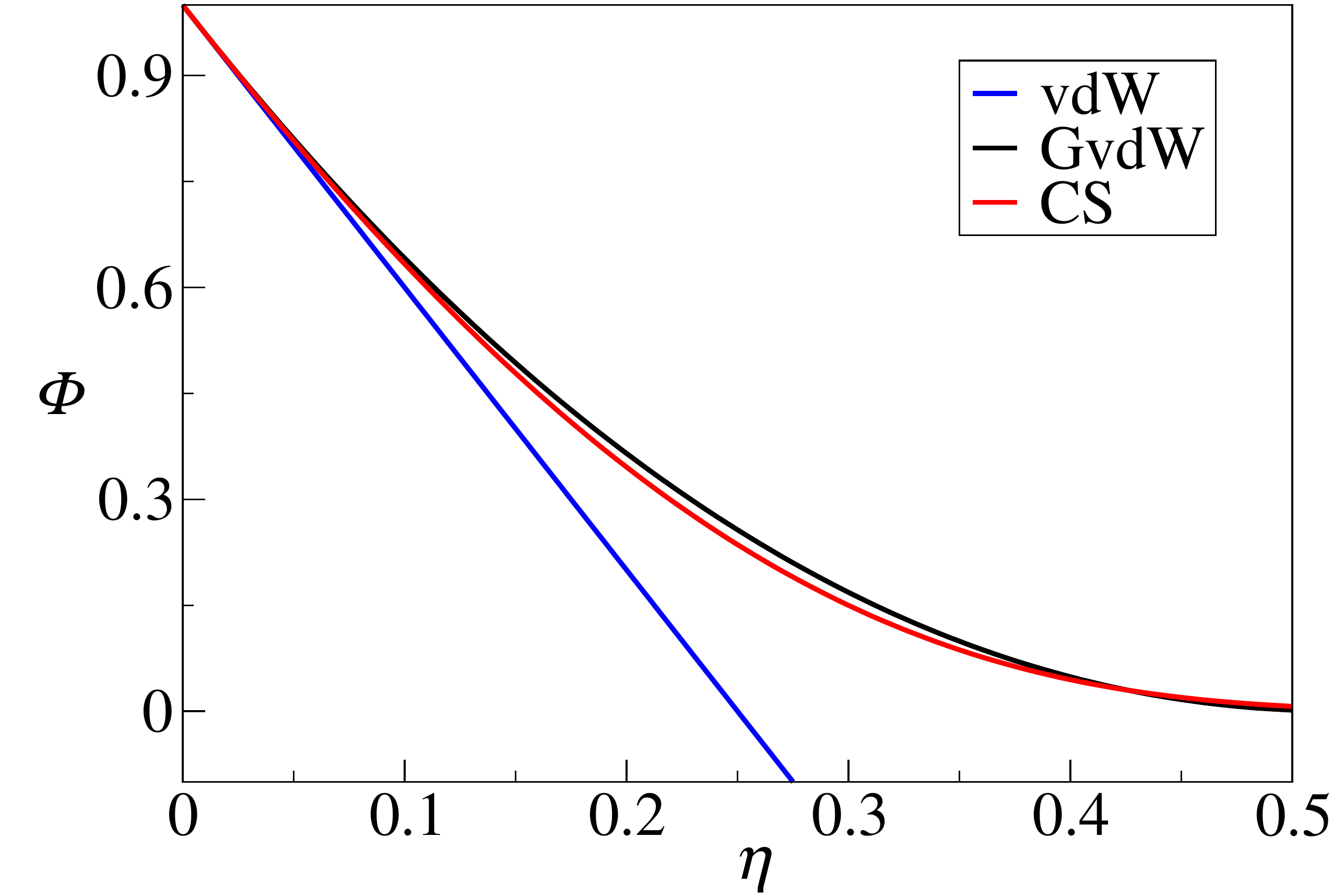}
\caption{Free volume fraction $\Phi$ as a function of the packing fraction 
	$\eta$ for a system of hard spheres. vdW: van der Waals. GvdW: 
	generalized van der Waals. CS: Carnahan-Starling. }
\label{FreeVol}
\end{figure}

However, it is perfectly possible to use another expression for 
$J$ which describes the reference system of HS. This suggests we 
can adopt an alternative PL scheme in order to study the NI 
transition at finite densities. An interesting proposal for 
the free energy $J$ is presented by Nordholm \textit{et al.} 
\cite{Nord} and Freasier \textit{et al.} \cite{Freasier} through 
a GvdW theory. According to this theory, the partition function 
is related to the free volume fraction $\Phi$ available to 
particles, which is estimated approximately. As a result, GvdW 
approach gives 
\begin{equation} \label{Felex2}
 J^{\text{GvdW}} \left( \eta \right) \equiv -\ln \Phi = -\frac{ 2 \pi }{3}\ln 
 \left( 1 - \frac{6\eta}{\pi} \right).
\end{equation}
For diluted systems, this expression is consistent with the simple vdW theory,
\begin{equation} \label{Felex2a}
 J^{\text{vdW}} \left( \eta \right) = -\ln \left( 1 - 4\eta \right).
\end{equation}
Despite the fact that the model presents a mean-field character, 
GvdW approximation agrees qualitatively very well with the free volume 
fraction associated with the CS equation of state, as shown in Fig.(\ref{FreeVol}).

Thus, we introduce a DFT considering that the excess free energy is given by 
Eq.\eqref{Felex1} in addition to the function $J \left( \eta \right)$ 
in Eq.\eqref{Felex2}. Clearly, by construction, we find a excess free energy which 
is consistent with Nordholm \textit{et al.} \cite{Nord} and 
Freasier \textit{et al.} \cite{Freasier} for a system of HS 
at finite packing fractions. Also, it is possible to check that, at very 
low densities, we obtain an expression which is in agreement with 
Onsager's model \cite{Onsager} for hard rods,
\begin{equation} \label{Onsager}
 F_{ex}^{On} = N\frac{\rho kT}{2}\int d\hat{\boldsymbol{a}} d\hat{\boldsymbol{a}}^{\prime} 
 f(\hat{\boldsymbol{a}})  f(\hat{\boldsymbol{a}}^{\prime}) V_{e}\left( \hat{\boldsymbol{a}},\hat{\boldsymbol{a}}^{\prime} \right).
\end{equation}

\section{Stationary distribution function for a system of hard spheroids} \label{StatDis}

The total free energy (\ref{Helden1}) is written as the sum of the ideal part 
\eqref{Felid1} and the excess part (\ref{Felex1}). Consequently, the PL theory leads 
to the dimensionless free energy density
\begin{equation} \label{Helden2}
 \begin{split}
 \mathcal{F} & \equiv \frac{F}{kTV} = \rho\ln \rho -\rho + \rho \left\{ 
  \int d\hat{\boldsymbol{a}} f(\hat{\boldsymbol{a}}) \ln f(\hat{\boldsymbol{a}}) + \right. \\
   & \qquad +\frac{J\left(\eta\right)}{8v_{0}} \int d\hat{\boldsymbol{a}} 
     d\hat{\boldsymbol{a}}^{\prime} f(\hat{\boldsymbol{a}}) f(\hat{\boldsymbol{a}}^{\prime}) 
     V_{e}\left( \hat{\boldsymbol{a}},\hat{\boldsymbol{a}}^{\prime} \right) + \\  
   & \left. \qquad \quad + \lambda\left[ \int d\hat{\boldsymbol{a}} f(\hat{\boldsymbol{a}}) -1 \right] \right\}.
  \end{split}
\end{equation}
Note that we introduce the Lagrange's multiplier $\lambda$ due to the normalization constraint
\begin{equation} \label{const}
  \int d\hat{\boldsymbol{a}} f(\hat{\boldsymbol{a}}) = 1.
\end{equation}
The free energy \eqref{Helden2} can be used to describe dense fluid phases 
in a PL treatment regardless the quantity $J$ for the HS reference system.

The equilibrium distribution $f_{eq}(\hat{\boldsymbol{a}})$ is the one which 
minimizes the free energy density. This means that we should find the stationary 
distribution associated with \eqref{Helden2}. The constant $\lambda$ is eliminated by 
using the normalization condition (\ref{const}). Then, we have a 
variational problem which leads to an Euler-Lagrange 
equation. As a result, the stationary distribution function may be written as
\begin{equation} \label{dist1}
 \begin{split}
 f\left( \hat{\boldsymbol{a}} \right) &= \frac{1}{Z} \exp 
  \left[ -\frac{J\left(\eta\right)}{4v_{0}} 
  \int d\hat{\boldsymbol{a}}^{\prime} f(\hat{\boldsymbol{a}}^{\prime})V_{e} 
  \left(\hat{\boldsymbol{a}},\hat{\boldsymbol{a}}^{\prime} \right)\right], \\
 Z &= \int d\hat{\boldsymbol{a}} \exp \left[ -\frac{J\left(\eta\right)}{4v_{0}} 
  \int d\hat{\boldsymbol{a}}^{\prime} f(\hat{\boldsymbol{a}}^{\prime})V_{e} 
  \left(\hat{\boldsymbol{a}},\hat{\boldsymbol{a}}^{\prime} \right)\right].
 \end{split}
\end{equation}

For a pair of hard spheroids, the excluded volume is a function of the relative orientation given by
$\cos^{-1}\left( \hat{\boldsymbol{a}} \cdot \hat{\boldsymbol{a}}^{\prime} \right)$. In practice, it 
is very difficult to write $V_{e}$ in a simple form \cite{IsiharaPiastra}. However, we consider an 
approximation by taking the Legendre expansion of the excluded volume up to the second order,
\begin{equation} \label{LegenExp}
  V_{e} \left(\hat{\boldsymbol{a}},\hat{\boldsymbol{a}}^{\prime} \right) = v_{0} \left[ a_{0} - 
  a_{2}P_{2}\left( \hat{\boldsymbol{a}} \cdot \hat{\boldsymbol{a}}^{\prime} 
 \right) \right], 
\end{equation}
where $P_{2}$ is the second Legendre polynomial. The coefficients $a_{0}$ and $a_{2}$ can be estimated, 
for example, by using pair configurations with known excluded volume expressions. Alternatively, we may 
determine the coefficient $a_{0}$ by taking the isotropic average of Eq.(\ref{LegenExp}),
\begin{equation} \label{coefexc1}
 \langle V_{e} \rangle _{iso} = v_{0}a_0  ,
\end{equation} 
which is proportional to the second virial coefficient in the isotropic phase. Also, 
for a pair of parallel spheroids, the excluded volume is $8v_{0}$, which leads to
\begin{equation} \label{coefexc2}
v_{0}a_2 = \langle V_{e} \rangle _{iso} - 8v_{0}.
\end{equation}
The isotropic average of $V_{e}$, for general convex bodies 
\cite{Kihara1953}, is given analytically by
\begin{equation} \label{isoexcvol}
  \langle V_{e} \rangle _{iso} = 2v_{0} + \frac{AM}{2\pi},
\end{equation}
where  $A$ and $M$ are the surface area and the mean curvature, respectively. All those shape measures 
can be written explicitly in terms of the spheroid axial ratio $\kappa=a/b$.

\section{Equations of states and phase behavior} \label{EoS}

It is straightforward to obtain expressions for the pressure $P$ and the chemical 
potential $\mu$ in a free energy density representation through the usual thermodynamic 
formalism \cite{Callen}. We also may simplify the study by  assuming that only uniaxial 
nematic phases are described by the distribution function $f\left( \hat{\boldsymbol{a}} \right)$. 
That assumption is reasonable because statistical models for systems of axially 
symmetric bodies only give rise to uniaxial mesophases in the absence of external 
fields \cite{Vertogen,deGennes}. 

Then, using the total free energy density (\ref{Helden2}) and the 
distribution function (\ref{dist1}), we can write the scaled pressure,
\begin{equation} \label{ESPress}
 \begin{split}
  \overline{P} \equiv \frac{Pv_{0}}{kT}  = \eta + \frac{\eta^2}{8}\frac{\partial J}
    {\partial\eta} \left( a_0 - a_2 S^2 \right) ,
 \end{split}
\end{equation}
and the scaled chemical potential,
\begin{equation} \label{ESmu}
 \begin{split}
  \overline{\mu} & \equiv \frac{\mu }{kT} +\ln v_0 =  \ln\eta +  
   \frac{1}{8}\left( \eta\frac{\partial J}{\partial \eta} - J  \right) \times \\ 
     & \qquad \times \left( a_0 - a_2 S^2 \right) - \ln Z,
 \end{split}
\end{equation}
where
\begin{equation} \label{Opeq}
 S = \langle P_{2}\left( \hat{\boldsymbol{a}} \cdot \hat{\boldsymbol{n}} \right) 
     \rangle=\int d \hat{\boldsymbol{a}}  f\left( \hat{\boldsymbol{a}} \right)
      P_{2}\left( \hat{\boldsymbol{a}} \cdot \hat{\boldsymbol{n}} \right),
\end{equation}
is the average of the second Legendre polynomial, and $\hat{\boldsymbol{n}}$ 
is the nematic director. For a given packing fraction $\eta$, Eq.(\ref{Opeq}) 
is a self-consistent equation for $S$, which may be solved numerically. After 
finding a solution, the pressure and the chemical potential are calculated 
and the thermodynamic stability of the system is determined. The equations of 
state are written in a way one can use any expression for the dimensionless 
free energy $J$. Clearly, in this paper, we focus on the expressions \eqref{CS} 
and \eqref{Felex2} for $J$.
\begin{figure} [h]
\centering
\includegraphics[scale=0.3]{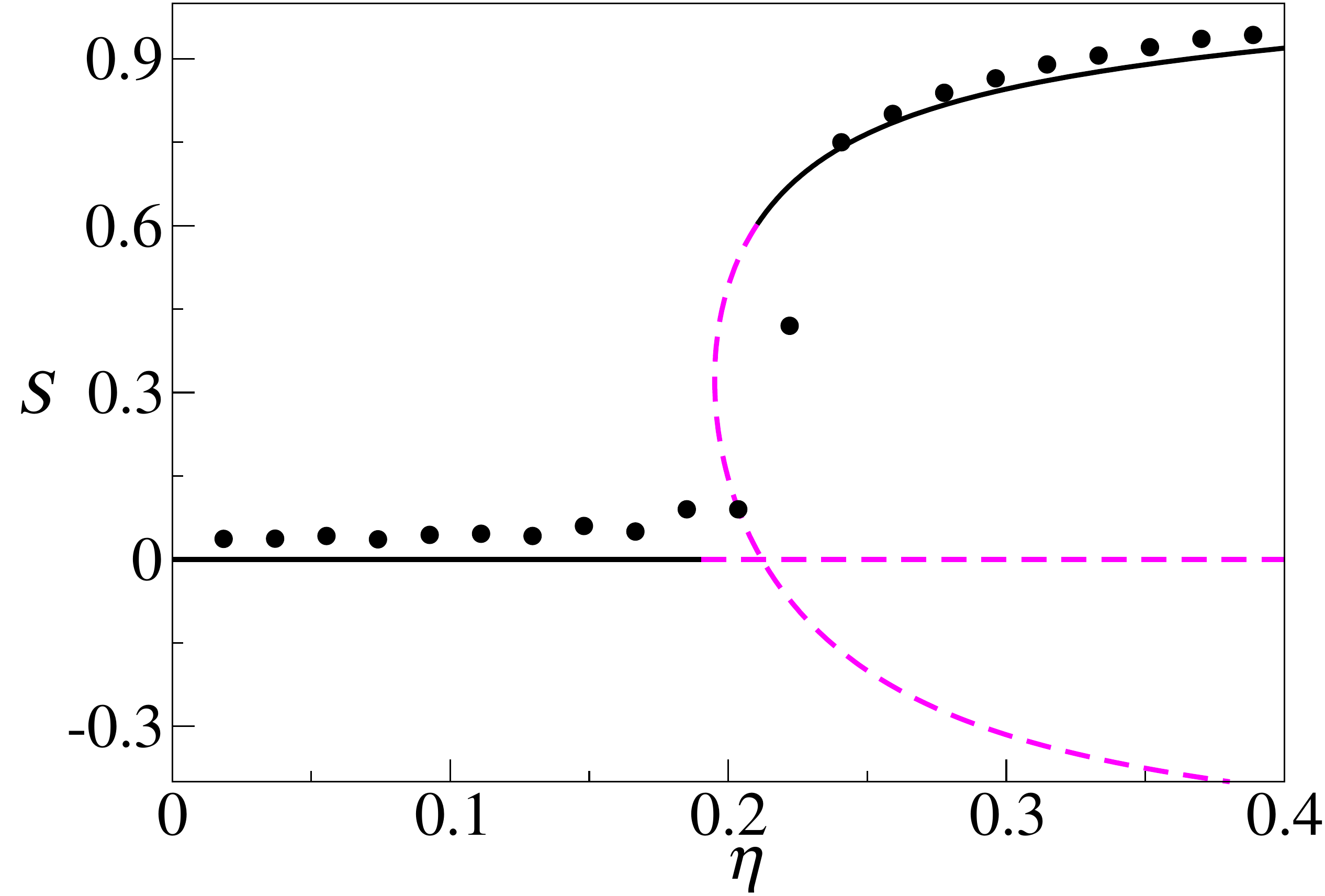}
\caption{ $S$ parameter versus packing fraction $\eta$ for axial ratio $\kappa = 10$. 
Black lines: stable solutions. Magenta dashed lines: non-stable solutions. Dots: MC 
simulations (A. Samborski \textit{et al.} \cite{Camp1996}).}
\label{OPk=10}
\end{figure}

Fig. \ref{OPk=10} shows the solutions of Eq.(\ref{Opeq}) for the axial ratio $\kappa=10$. 
There are two stable branches (black lines), one associated with the isotropic phase 
($S=0$) at low packing fractions, and other one related to a calamitic nematic phase 
($S > 0$) at high values of $\eta$. There are also non-stable (metastable and  unstable) 
solutions represented by magenta dashed lines. The isotropic branch becomes metastable 
as the packing fraction increases. A discotic phase ($S<0$) is presented at high $\eta$, 
but it is non-stable. In fact, calamitic states are the only nontrivial solutions 
thermodynamically favorable. The order parameter behavior is in good agreement with 
the MC data results presented by Camp \textit{et al.} 
\cite{Camp1996}.

\begin{figure} [h]
\centering
\includegraphics[scale=0.3]{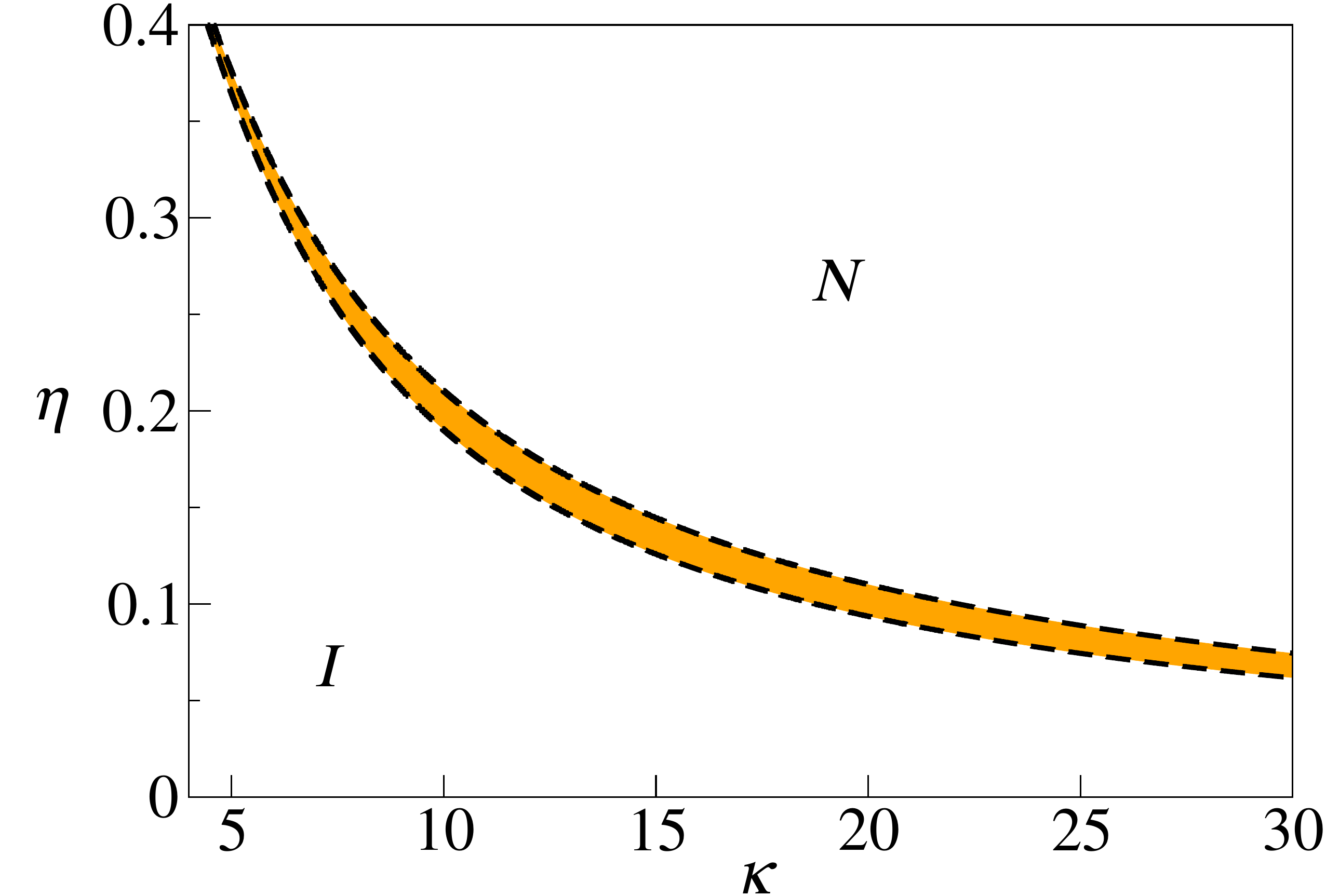}
\caption{\label{pc} Packing fraction $\eta$ versus axial ratio $\kappa$ for a system 
 of hard spheroids. $I$: isotropic. $N$: nematic. Orange region: NI phase coexistence.}
\end{figure}

The NI transition is discontinuous, with a density jump at the phase coexistence, 
as depicted in Fig. \ref{pc}. Given a particular value of axial ratio $\kappa$, the 
system goes from an isotropic phase $I$ to a nematic phase $N$ as the packing fraction 
increases. For very long spheroids ($\kappa \gg 1$), we recover Onsager's limit, 
characterized by long rodlike shapes and low densities. Also, it is possible to 
identify a Maxwell construction in a pressure versus volume graph. In fact, it is 
straightforward to show that the Maxwell construction is equivalent to the 
equilibrium conditions that characterizes the NI phase separation.

\begin{figure}
\centering
\includegraphics[scale=0.3]{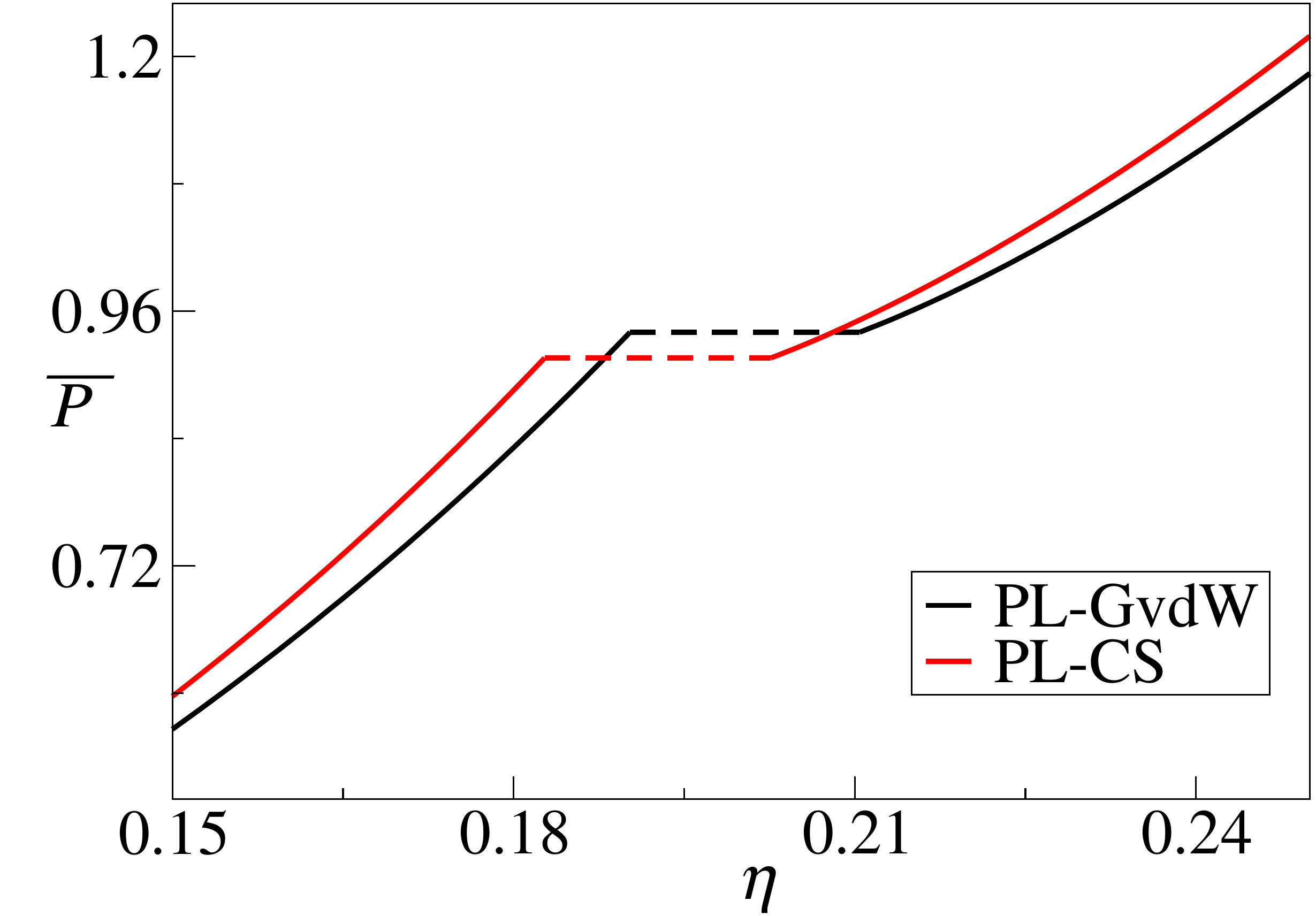}
\caption{Scaled pressure $\overline{P}$ versus packing 
	fraction $\eta$ for axial ratio $\kappa = 10$. PL-GvdW: Parsons-Lee theory with 
	generalized van der Waals approximation. PL-GvdW: Parsons-Lee theory with 
	Carnahan-Starling approach.}
\label{PressEoS}
\end{figure}

The usual PL theory adopts a CS equation of state for the hard-sphere reference system 
(PL-CS). The approach we propose here follows the main ideas of the PL scheme, but the 
system of hard spheres is described by an generalized van der Waals approximation (PL-GvdW). 
It is important to compare both models, since we introduce an alternative approach to 
the NI transition. We can do that by studying the equation of state in the vicinity of 
the phase transition. For example, the equilibrium pressure behavior is presented in 
Fig. \ref{PressEoS} for axial ratio $\kappa=10$. The PL-GvdW leads to a pressure 
transition higher than PL-CS results. However, both treatments seem to agree 
qualitatively well for the given axial ratio.  

\begin{figure} [h]
\centering
\includegraphics[scale=0.3]{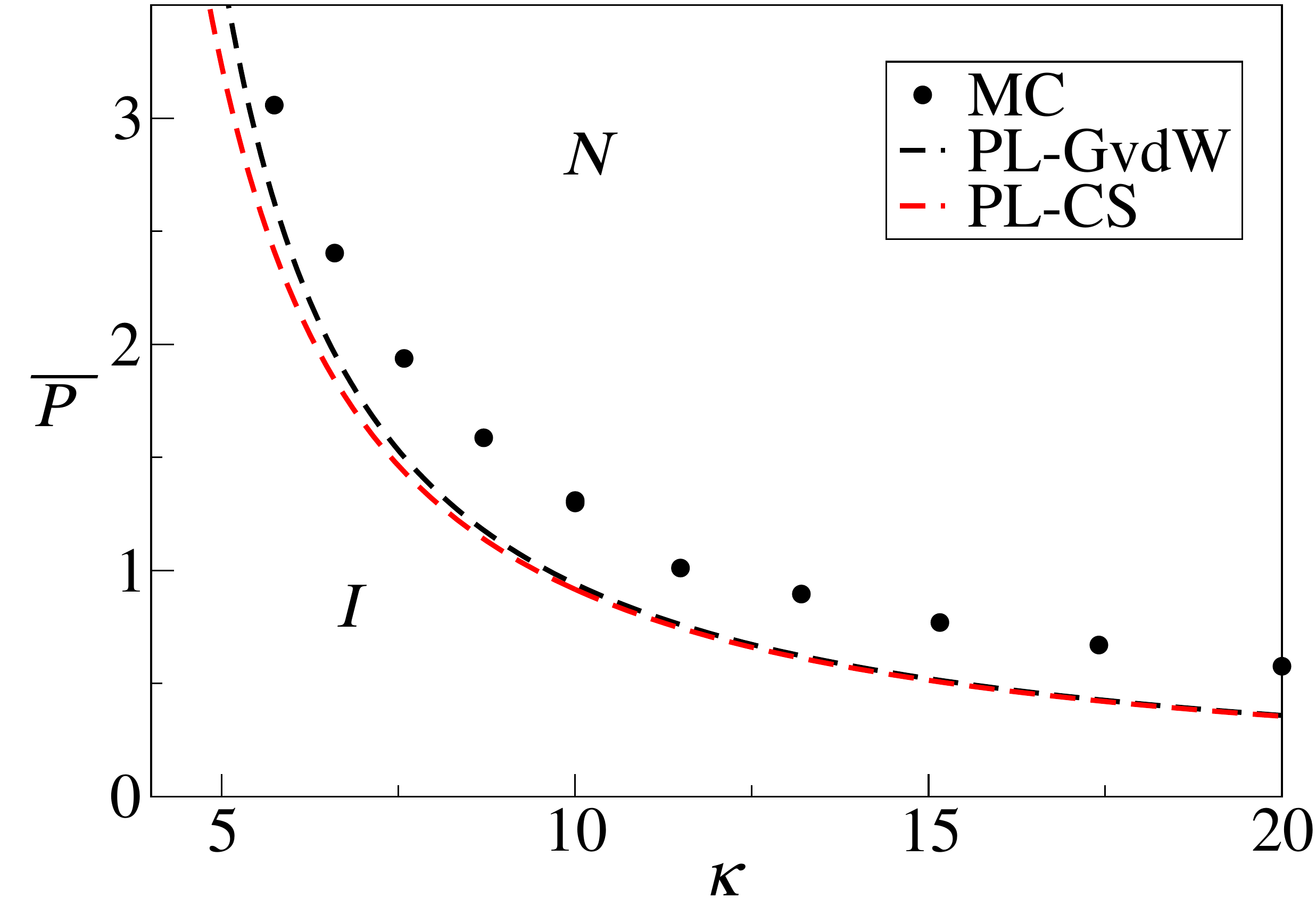}
\caption{\label{McPr}  Scaled pressure $\overline{P}$ versus axial ratio 
 $\kappa$ phase diagram. Black dashed lines: PL with GvdW. Red dashed lines: 
 PL with CS. Dots:  MC results (Camp \textit{et al.}\cite{Camp1996}). }
\end{figure}

Also, it should be interesting to compare the results with approaches which consider 
fluctuations more appropriately, since we have a theory of mean-field type. 
Fig. \ref{McPr} shows the $\overline{P} \times \kappa$ phase diagram according 
to the DFT presented in this work, the usual PL-CS theory and the MC simulations 
from Camp \textit{et al.}\cite{Camp1996}. Our mean-field model underestimates 
the phase boundary for the packing fraction range considered compared with MC 
data. However, PL-GvdW leads to a phase boundary which agrees very well with 
PL-CS. This surprisingly result clearly shows the robustness of the decoupling 
approximation which characterizes the PL theory. 

The CS equation of state is one of the best approximations for studying HS systems. 
However, the GvdW scheme, based on heuristic arguments, also leads to reasonable 
good results. Using a GvdW approximation, we have a DFT which describes 
qualitatively well the phase behavior of anisotropic fluids at finite densities. 
Then, despite the simplicity, we argue that PL theory combining with GvdW 
approach can be useful to study hard-core effects in systems 
of convex non-spherical bodies.

\section{Conclusions} \label{Con}

We present a DFT constructed by using the Parsons-Lee theory and a generalized van der 
Waals approximation in order to investigate systems of hard anisotropic bodies. For 
the case of spheroidal shapes, the pair excluded volume is approximated to a truncated 
Legendre expansion, which simplifies the analysis. The model exhibits a NI discontinuous 
transition characterized by a phase coexistence. The isotropic phase is stable at low 
densities and the nematic phase becomes thermodynamically favorable as the packing fraction 
increases. The theory agrees qualitatively with MC results for finite axial ratios. Then, 
we believe the model is appropriate to study the macroscopic properties of non-spherical 
particle systems with steric interactions. Further investigations may consider the inclusion 
of attractive pair potentials, which may lead to a kind of van der Waals equation of state 
for a system of anisotropic molecules, as well to a richer phase behavior. From the 
statistical mechanics point of view, the model can also be extended to include disordered 
degrees of freedom in order to mimic shape variation effects, for example, in multicomponent 
mixtures.

\section*{Acknowledgement}
We thank the anonymous referees for their insightful comments 
and suggestions. The author is indebted to Peter Palffy-Muhoray 
for the stimulating visiting period and the kind hospitality of the 
Liquid Crystal Institute. Also, the author is grateful to A. P. Vieira and 
S. R. Salinas for critically reading the manuscript. This work was 
supported by the grants 2016/07448-5 and 2013/12873-9, São Paulo 
Research Foundation (FAPESP). The final publication is available at Springer
via \url{http://dx.doi.org/10.1140/epje/i2018-11746-0}.

\end{document}